\documentclass[a4paper,pra,reprint, twocolumn,superscriptaddress]{revtex4}

\usepackage{ulem}
\usepackage{amssymb}
\usepackage{amsmath}
\usepackage{epsfig}
\usepackage{color}
\usepackage{graphics, graphicx}
\usepackage{bbold}
\usepackage{psfrag}
\usepackage{mathcomp}
\usepackage{amsmath}
\usepackage{amssymb}%花体字母加粗
\usepackage{mathrsfs}%花体字母
\usepackage{subfigure}
\usepackage{verbatim}
\usepackage{xcolor}
\usepackage[colorlinks,citecolor=blue,urlcolor=blue]{hyperref}

\usepackage{mathrsfs}

\begin{document}

\date{\today}
\title{Chiral Majorana edge states in the vortex core of a $p+ip$ Fermi superfluid}
\author{Jing-Bo Wang}
\affiliation{Key Laboratory of Quantum Information, Chinese Academy of Sciences, School of Physics, University of Science and Technology of China, Hefei, Anhui, 230026, China}
\affiliation{Synergetic Innovation Center of Quantum Information and Quantum Physics, University of Science and Technology of China, Hefei, Anhui 230026, China}
\author{Wei Yi}
\email{wyiz@ustc.edu.cn}
\affiliation{Key Laboratory of Quantum Information, Chinese Academy of Sciences, School of Physics, University of Science and Technology of China, Hefei, Anhui, 230026, China}
\affiliation{Synergetic Innovation Center of Quantum Information and Quantum Physics, University of Science and Technology of China, Hefei, Anhui 230026, China}
\author{Jian-Song Pan}
\email{panjsong@sjtu.edu.cn}
\affiliation{Wilczek Quantum Center, School of Physics and Astronomy and T. D. Lee Institute,
Shanghai Jiao Tong University, Shanghai 200240, China}
%\affiliation{Synergetic Innovation Center of Quantum Information and Quantum Physics, University of Science and Technology of China, Hefei, Anhui 230026, China}

\begin{abstract}
We study a single vortex in a two-dimensional $p+ip$ Fermi superfluid interacting with a Bose-Einstein condensate. The Fermi superfluid is topologically non-trivial and hosts a zero-energy Majorana bound state at the vortex core. Assuming a repulsive $s$-wave contact interaction between fermions and bosons, we find that fermions are depleted from the vortex core when the bosonic density becomes sufficiently large. In this case, a dynamically-driven local interface emerges between fermions and bosons, along which chiral Majorana edge states should appear.
We examine in detail the variation of vortex-core structures as well as the formation of chiral Majorana edge states with increasing bosonic density. In particular, when the angular momentum of the vortex matches the chirality of the Fermi superfluid, the Majorana zero mode and normal bound states within the core continuously evolve into chiral Majorana edge states. Otherwise, a first-order transition occurs in the lowest excited state within the core, due to the competition between counter-rotating normal bound states in forming chiral Majorana edge states. Such a transition is manifested as a sharp peak in the excitation gap above the Majorana zero mode, at which point the Majorana zero mode is protected by a large excitation gap.
Our study presents an illuminating example on how topological defects can be dynamically controlled in the context of cold atomic gases.
\end{abstract}

\maketitle

\section{Introduction}
Topological edge states emerge at the interface between phases of distinct topological nature, whose characterization and manipulation are among central issues in the study of topological materials. As an outstanding example, edge modes at vortex cores of spinless $p+ip$ superfluids or $\nu=5/2$ fractional quantum-Hall systems are non-Abelian Majorana zero modes (MZMs)~\cite{Moore1991Nonabelians, Read2000Paired}. Motivated by the pursuit of novel fundamental physics and potential applications in quantum computation, much effort has been devoted to the study of these non-Abelian MZMs in condensed-matter~\cite{Fu2008Superconductor, Sun2016Majorana, He2017Chiral,nayak2008nonabelian} or cold-atoms systems~\cite{Gurarie2005Quantum, Tewari2007Quantum}.

Generically, the existence and number of topological edge modes are dictated by topological invariants through the bulk-boundary correspondence~\cite{hasan2010colloquium,qi2011topological,stone2004edgemode}. However, properties of edge modes at a given boundary are affected by the geometry of the boundary. For example, a vortex in a two-dimensional topological $p+ip$ Fermi superfluid is a point defect, and the edge mode associated with such a point defect is an MZM bound to the core. By contrast, for a linear defect, the corresponding topological edge modes are linearly dispersive and reside on the one-dimensional boundary~\cite{gurarie2007zero,tsutsumi2008majorana,mizushima2008role,akhmerov2009electrically}.
A subtle scenario arises in two dimensions when a local defect develops a finite spatial expanse but is still localized within a closed contour. Here, while the existence of topological edge modes are still associated with bulk topological invariants, their forms of existence are closely related to the spatial geometry and range of the defect. Understanding the response of topological edge modes to geometrical deformations of defects provides valuable knowledge necessary for the manipulation and control of edge states.

In this work, we study the response of topological edge modes to dynamically-controlled defects in a mixture of Bose and Fermi condensates in two dimensions.
In a previous study~\cite{Pan2017Vortex}, it has been shown that, for a Bose-Fermi superfluid mixture in three dimensions, Bose-Einstein condensate (BEC) can become localized at the vortex core of the Fermi superfluid at a sufficiently large repulsive inter-species interaction energy. The localization of BEC thus dynamically generates an interface between fermions and bosons.

Building upon such a physical picture, we consider a single vortex in a $p+ip$ Fermi superfluid and in the presence of an atomic BEC, where fermions interact repulsively with atoms in the BEC.
The $p+ip$ superfluid is topologically nontrivial in the weak-pairing BCS regime, where its chemical potential $\mu_F>0$, and becomes topologically trivial in the strong-pairing BEC regime with $\mu_F<0$~\cite{Read2000Paired,gurarie2005quantumphase}. In this work, we focus on the case of a topological Fermi superfluid.
In the absence of BEC, vortices in the topological superfluid hosts MZMs at their centers~\cite{matsumote2001vortex,fendly2007edgestate,tewari2007index,moller2011structure,murray2015majorana,yaacov2009majorana,mizushima2010vortex}. These MZMs obey non-Abelian statistics and are robust against symmetry-preserving perturbations~\cite{Tewari2007Quantum,ivanov2001nonabelian,tsutsumi2015symmetry}. We show that, when the density of BEC is sufficiently large, the vortex-core structure is drastically modified as fermions are depleted from the core. This gives rise to a dynamically-generated boundary inside the vortex core. We find that the MZM and discrete Caroli-de Gennes-Matricon (CdGM) states in the core continuously merge into a branch of chiral Majorana edge states, provided the angular momentum of the vortex matches the chirality of the Fermi superfluid.
By contrast, a first-order transition occurs in the first excited state within the vortex core, where CdGM states with angular momenta of opposite signs compete with each other.
Our results present an interesting example where topological defects and topological edge modes are dynamically controlled.

While $p+ip$ Fermi superfluid may be experimentally implemented in a quasi-two-dimensional fully-polarized Fermi gas close to a $p$-wave Feshbach resonance, topological superfluid can also be realized in a two-dimensional two-component Fermi gas with strong $s$-wave interactions under two-dimensional spin-orbit coupling and out-of-plane Zeeman field~\cite{yi2015pairing,hu2013universal,liu2012probing, zhang2013topological}, where similar phenomena are expected in the presence of BEC.
Following the recent experimental realization of two-dimensional spin-orbit coupling in optical lattices~\cite{Wu83realization} and Fermi-Bose superfluid mixtures~\cite{Ferrier2014mixture,roy2017twomixture,yao2016observation,desalvo2017observation}, we hope that the vortex-core chiral edge states discussed in this work can be probed in the future.

This paper is organized as follows. In Sec. II, we present a brief derivation of the Bogoliubov-de-Gennes (BdG) equations describing the $p+ip$ superfluid, which is coupled with the Gross-Pitaevski (GP) equation describing the BEC. In Sec. III, we study the behavior of a single vortex with a matching angular momentum for the Fermi-superfluid chirality, and we discuss the opposite case in Sec. IV. Finally, we summarize in Sec. V.

\begin{figure}[tbp]
  \centering
  \includegraphics[width=7cm]{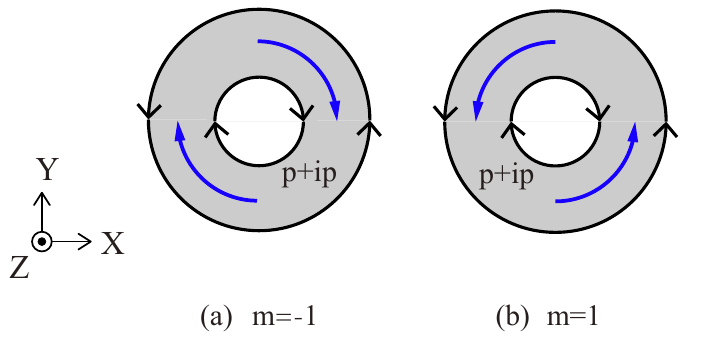}
  \caption{Schematic illustration of the chirality of Majorana edge states at the inner and outer boundaries of an annulus. In the context of our work, the inner boundary represents the dynamically-generated interface at the vortex core, and the outer boundary corresponds to the enforced open boundary. (a) When the angular momentum of the vortex is $m=-1$, the running direction of chiral Majorana edge states on the inner boundary (black arrows) matches the direction of circulation of the vortex (blue arrows). (b)  When the angular momentum of the vortex is $m=1$, the running direction of chiral Majorana edge states on the inner boundary is opposite to the direction of circulation of the vortex.}
  \label{fig:fig1}
\end{figure}

%problems:
%dimensionless units
%justification for gp in (quasi-)two dimensions
%BEC with vortex as a more exotic way of engineering dynamic interface

\section{Model}
We study the mixture of a $p+ip$ Fermi superfluid and a BEC by numerically solving the coupled BdG and GP equations.
Since the $p+ip$ Fermi pairing breaks the time-reversal symmetry, the BdG equation has only the particle-hole symmetry, such that the resulting topological superfluid belongs to the D class of the $Z$ classification~\cite{Teo2010Topological}. This means that a system with an open boundary also possesses a chiral edge mode even in the absence of vortices. The introduction of a vortex into the $p+ip$ superfluid should reduce the classifying group to $Z_2$. Hence, as long as the angular-momentum quantum number $m$ of the vortex is odd, an MZM should appear in the vortex core. In the current case, we consider a single vortex with $m=\pm 1$.
While the chirality of the Fermi superfluid is fixed, we will show that the vortex-core structures can be quite different for $m=1$ and $m=-1$, particularly in the presence of BEC. These two scenarios are schematically illustrations in Fig.~\ref{fig:fig1}.

\begin{figure}[tbp!]
  \centering
  \includegraphics[width=8cm]{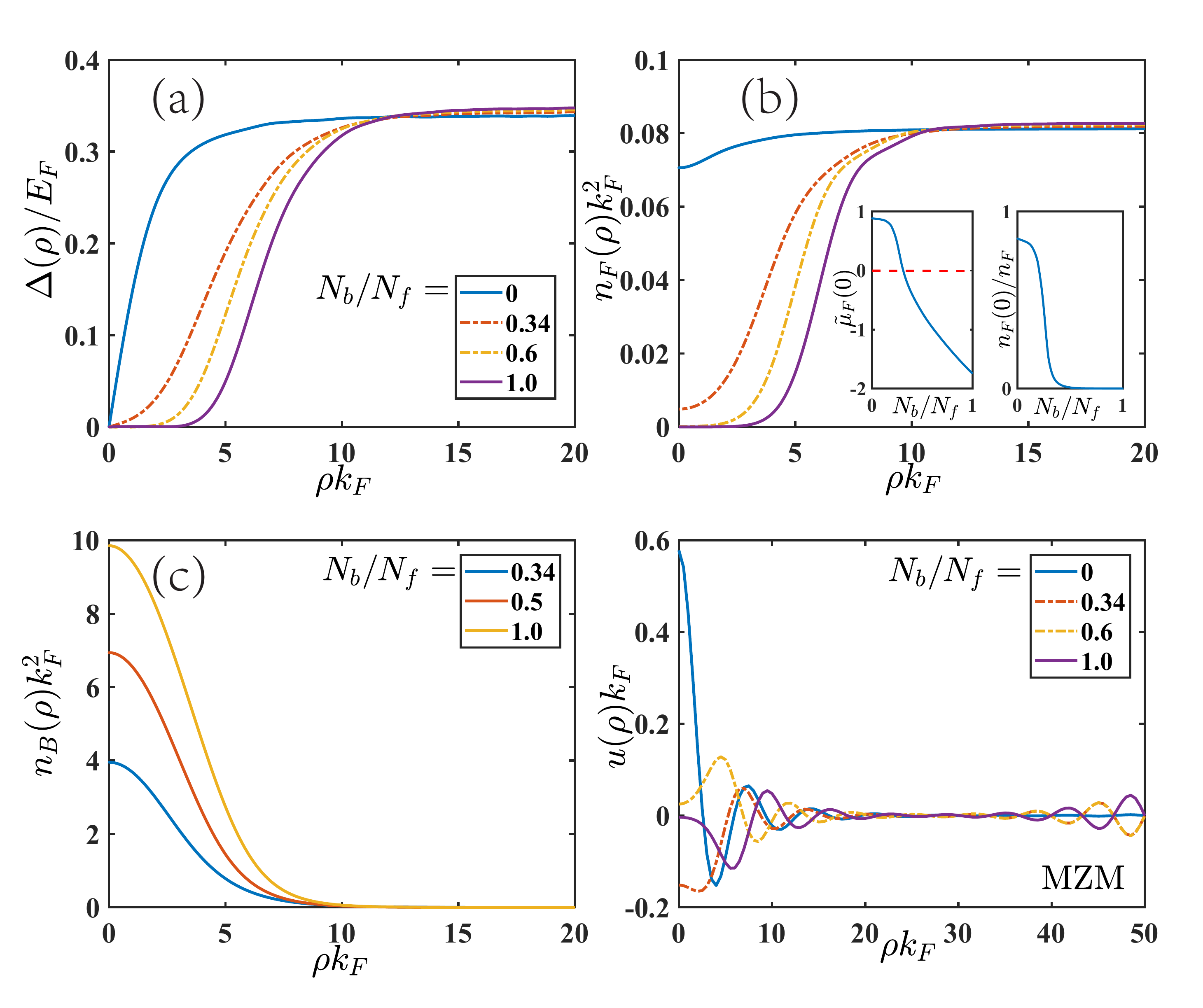}
  \caption{Spatial profiles of various quantities of a vortex core with $m=-1$ and
  for different atomic-number ratios between the bosonic and fermionic components $N_b/N_f$. (a) The pairing order parameter $\Delta(\rho)$, (b) the Fermi density distribution $n_F(\rho)$, (c) the bosonic density distribution $n_B(\rho)$, and (d) the wave function of the MZM $u(\rho)$. Note that for MZM $v(\rho)=u(\rho)$. Insets in (b): variations of the effective chemical potential $\tilde{\mu}_{F}(0)=\mu_F-g_{BF}n_B(0)$ (left) and the Fermi density $n_F(0)$ (right) at the center of the vortex core with increasing $N_b/N_f$. For simplicity, the Bose-Bose interaction rate $g_B$ is set to be zero. We use the Fermi energy $E_F=k_F^2/2m_F$ as the unit of energy. For numerical calculations, we take $R=50k_F^{-1}$ as the outer boundary, and we fix fermion atom number $N_f=604$. The scattering length for the Bose-Fermi interaction is $80 a_0$ with $a_0$ the Bohr radius, and the Fermi-Fermi interaction is tuned to the BCS regime with $\mu_F=0.9$.}
  \label{fig:fig2}
\end{figure}

\begin{figure*}[tbp!]
  \centering
  \includegraphics[width=13cm]{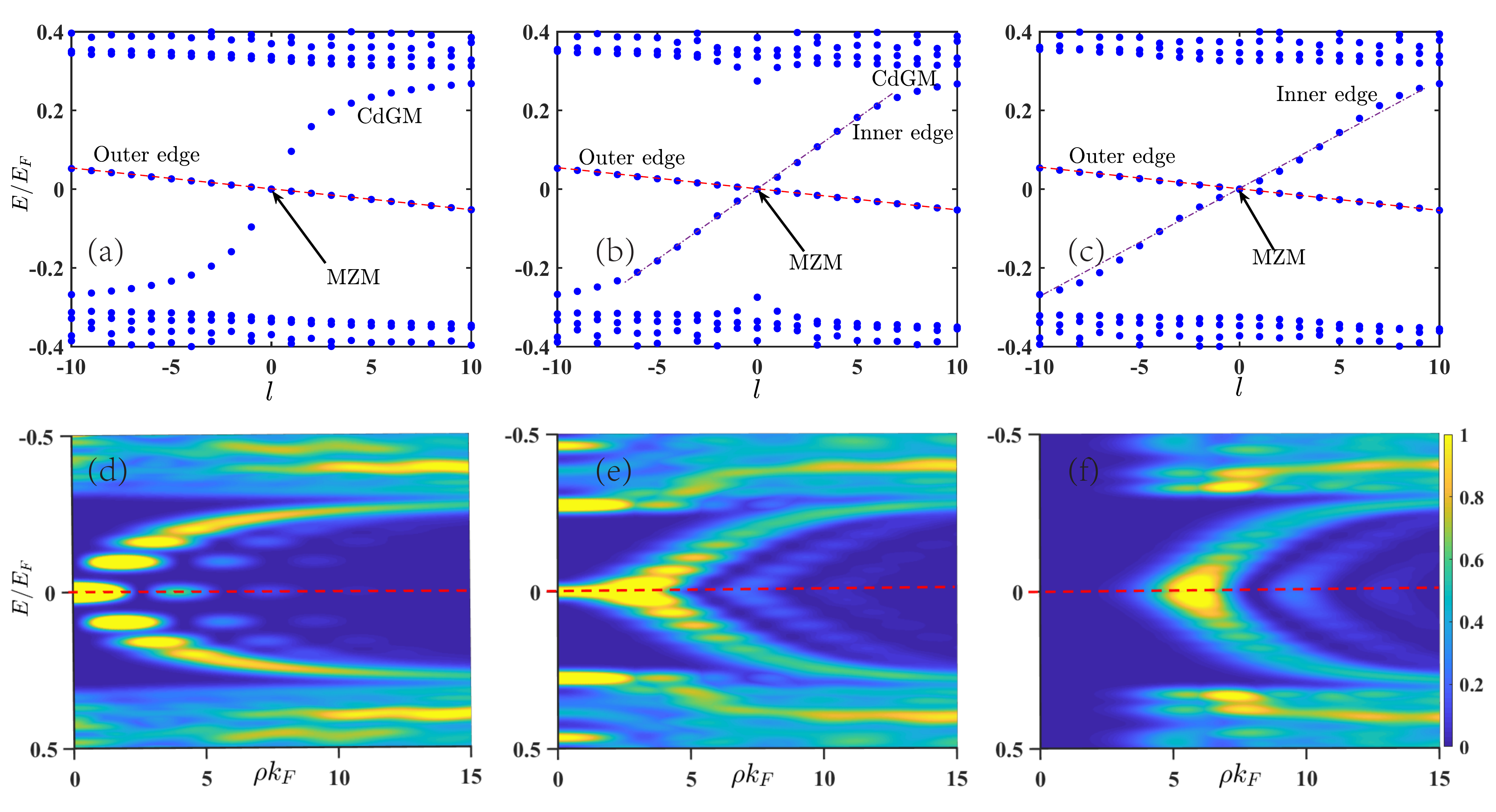}
  \caption{Quasiparticle spectra (a)(b)(c) and LDOS $D(\rho,E)$ (d)(e)(f) of a single vortex with $m=-1$. (a)(d)  In the absence of BEC with $N_b/N_f=0$; (b)(e)$N_b/N_f=0.34$; (c)(f) $N_b/N_f=1$ with a fully depleted vortex core. All other parameters are the same with those in Fig.~\ref{fig:fig2}.}
  \label{fig:fig3}
\end{figure*}

The BdG equation of the fermion quasiparticle wave function ${[u_n,v_n]}^T$ is given by
\begin{align}
\left[\begin{matrix} H_0(\mathbf{r})&\Pi(\mathbf{r})\\ -\Pi^\ast(\mathbf{r})&-H_0(\mathbf{r})\end{matrix}\right]\left[\begin{matrix} u_n(\mathbf{r})\\ v_n(\mathbf{r})\end{matrix}\right]=E_n\left[\begin{matrix} u_n(\mathbf{r})\\ v_n(\mathbf{r})\end{matrix}\right],
\label{equation:BdG2}
\end{align}
where the off-diagonal term is defined as $\Pi(\mathbf{r})=\frac{i}{k_F}\left[\Delta(\mathbf{r})P+\frac{1}{2}P\Delta({\mathbf{r}})\right]$ with the operate $P=\partial_x+i\partial_y$. For convenience, we set $\hbar=1$ throughout this work. $\Delta(\mathbf{r})$ is the $p$-wave pairing order parameter of the $p$-wave Fermi superfluid. Here, the diagonal term $H_0(\mathbf{r})=-\frac{\nabla^2}{2m_F}-\tilde{\mu}_F(\boldsymbol{r})$, with the effective Fermi chemical potential $\tilde{\mu}_F(\boldsymbol{r})=\mu_F-g_{BF}|\phi(\mathbf{r})|^2$, the $s$-wave Fermi-Bose interaction $g_{BF}=\frac{2\pi(m_B+m_F)a_{BF}}{m_Bm_F}$, and the bosonic ground-state wave function $\phi(\mathbf{r})$.

The dynamics of the BEC is described by the GP equation
\begin{align}
\Big[-\frac{\nabla^2}{2m_B}+g_{BF}n_F(\mathbf{r})+g_Bn_B(\mathbf{r})\Big]\phi(\mathbf{r})=\mu_B\phi(\mathbf{r}),
\end{align}
where $\mu_B$ is the bosonic chemical potential, $g_B=\frac{4\pi a_B}{m_B}$ is the boson-boson interaction coefficient, $n_F(r)=\sum_{E_n>0}(|u_n|^2+|v_n|^2)$ and $n_B(r)=|\phi(\mathbf{r})|^2$ are the fermion and boson densities, respectively.

The $p+ip$ Fermi order parameter is
\begin{equation}
 \Delta(\mathbf{r})=\frac{ig_p}{k_F}\sum_{E_n>0}\Big[v^\ast_n(\mathbf{r})(\partial_x-i\partial_y)
u_n(\mathbf{r})-u_n\leftrightarrow v_n\Big],
\end{equation}
where $g_p=\int_0^{k_c}\frac{d\vec{k}}{(2\pi)^2}\frac{(k/k_F)^2}{k^2/m_F-E_b}$ is the effective $p$-wave interaction strength and $k_F=\sqrt{4 N_f/R^2}$ with the radius of the open boundary $R$, and the Fermi particle number $N_f=2\pi \int_{0}^{R} dr r n_F(r)$ is the Fermi momentum. $E_b$ is the two-body bound-state energy of fermions in vacuum and $k_c$ is the momentum cutoff, which we set through $k_c^2/2m_F=10E_F$ in our calculation with the Fermi energy $E_F=k^2_F/2m_F$.

To make the coupled BdG and GP equations consistent, we rewrite the order parameter as $\Delta(\mathbf{r})=\Delta(\rho)e^{im\theta}$ in the polar coordinate. As discussed earlier, we mainly focus on the case where $m=\pm 1$. Under an open-boundary condition with rotational symmetry, we expand the radial component of the particle and hole wave functions under the Fourier-Bessel basis as
\begin{align}
\left[\begin{matrix} u_{n}(\rho)\\v_{n}(\rho)\end{matrix}\right]
=\sum_{j,l}\left[\begin{matrix} c_{n,l}^{(j)}\varphi_{j,l}(\rho)\\d_{n,l}^{(j)}\varphi_{j,l-m-1}(\rho)\end{matrix}\right],
\end{align}
where $\varphi_{jl}(\rho)=\frac{\sqrt 2J_l(\alpha_{jl}\rho/R)}{RJ_{l+1}(\alpha_{jl})}$, $J_l(r)$ is the $l$-th order Bessel function, $\alpha_{jl}$ is the $j$-th root of $J_l(r)$.
The BdG equation then becomes
\begin{align}
\sum_{j'}\left[\begin{matrix} T_{l}^{jj'}&\Delta_{l}^{jj'}\\ \Delta_{l}^{jj'}&-T_{l-m-1}^{jj'}\end{matrix}\right]
\left[\begin{matrix} c_{n,l}^{(j')}\\ d_{n,l}^{(j')}\end{matrix}\right]
=E_n\left[\begin{matrix} c_{n,l}^{(j)}\\ d_{n,l}^{(j)}\end{matrix}\right]
\end{align}
where
\begin{align}
T_{l}^{jj'}=&\Big(\frac{1}{2m_F}\frac{\alpha^2_{j,l}}{R^2}-\mu_F\Big)\delta_{jj'}
+g_{BF}n_{B,l}^{jj'},\\
n_{B,l}^{jj'}=&\int_0^R\rho d\rho n_B(\rho)\varphi_{j,l}(\rho)\varphi_{j',l-m-1}(\rho),\\
\Delta_{l}^{jj'}=&\int_0^R\rho d\rho\Big(\Delta(\rho)\chi_{j',l-m-1}\varphi_{j,l}(\rho)\Big)\nonumber\\
&-\frac{1}{2}\int_0^R\rho d\rho\Big(\frac{\partial\Delta(\rho)}{\partial\rho}-\frac{m\Delta(\rho)}{\rho}\Big)
\varphi_{j',l-m-1}(\rho)\varphi_{j,l}(\rho)\nonumber,
\end{align}
and
\begin{equation}
\Delta(\rho)=\frac{g_p}{2\pi}\sum_{l,E_n\geq0,jj'}c_{j'l}^{n}d_{jl}^{n}
\Big(\varphi_{j,l}\chi^\ast_{j',l-m-1}+\varphi_{j',l-m-1}\chi_{j,l}\Big).
\end{equation}
Here we define $\chi_{jl}(\rho)=\frac{\alpha_{jl}}{R}\frac{\sqrt 2J_{l+1}(\alpha_{jl}\rho/R)}{RJ_{l+1}(\alpha_{jl})},
\chi^\ast_{jl}(\rho)=\frac{\alpha_{jl}}{R}\frac{\sqrt 2J_{l-1}(\alpha_{jl}\rho/R)}{RJ_{l+1}(\alpha_{jl})}$.
By self-consistently solving the coupled BdG and GP equations under the open boundary condition, we obtain wave functions of the Fermi quasi-particles and the stable ground state of BEC. For convenience, we only consider the case where no external trapping potentials are present.

\section{Chiral Majorana edge states in a vortex with $m=-1$}
In this section, we study the vortex-core structure for the case of $m=-1$.
In Fig.~\ref{fig:fig2}, we show typical quantities characterizing the Fermi vortex-core structure.
In the absence of bosons, the order parameter $\Delta(\rho)\propto\rho^{|m|}$ when $\rho$ is small. As the Bose-Fermi interaction is repulsive, the depletion of fermions in the vortex core effectively attracts bosons. The condensation of bosons at the vortex core should then further enhance the fermion depletion~\cite{Pan2017Vortex}. As shown in Fig.~\ref{fig:fig2}(a)(b)(c), the vortex core is completely depleted above a threshold BEC density. This is driven by a negative effective Fermi chemical potential $\tilde{\mu}_{F}(0)=\mu_F-g_{BF}n_B(0)<0$ near the vortex core [see the left inset in Fig~\ref{fig:fig2}(b)].
We note that a complete depletion of the vortex core occurs more favorably at small or negative Bose-Bose interaction $g_B$, since in this case the BEC can easily condense into the vortex core, either due to a small coherence length $\xi_B$ or due to the attractive interaction. For simplicity, we set the Bose-Bose interaction to zero for our numerical calculations here.

Interestingly, as shown in Fig.~\ref{fig:fig2}(d), the spatial wave function of the MZM is deformed into a ring when the vortex core is completely depleted. The increase in the vortex size should make both vortices and MZMs more accessible in experiments.
The condensation of bosons in the vortex core further provides the possibility of adiabatically controlling the trajectory of the vortex by spatially shifting the BEC by tuning its trapping lasers. Such a possibility can be useful in future applications like topological quantum computation based on topological superfluid, where the manipulation of qubits can be implemented by braiding vortices with MZMs.

To further characterize vortex-core structures and MZMs in the presence of BEC, we calculate the quasiparticle spectra and local density of states (LDOS) of the system. Here, the LDOS can be written as $D(r,E)=\sum_{n} \left[|u_{n}(r)|^2\delta(E-E_{n})+|v_{n}(r)|^2\delta(E+E_{n})\right]$, which yields the distribution of eigenstates with energy $E$ in the position space.
In Fig.~\ref{fig:fig3}, we plot quasiparticle spectra and the corresponding LDOS for cases without (a) and with (b)(c) BEC, respectively.
As shown in Fig.~\ref{fig:fig3}(a), in the absence of BEC, the branch of chiral Majorana edge states at the outer boundary $R$ crosses the zero-energy point, which gives rise to a two-fold degeneracy at zero energy. The MZM at the vortex core is manifested as
the zero-energy bright horizontal stripe near $\rho=0$ in Fig.~\ref{fig:fig3}(d). In contrast, the discrete CdGM modes, arising from the Andreev reflection, all have finite energies [see Fig.~\ref{fig:fig3}(a)(d)]~\cite{caroli1964bound,ioselevich2013tunneling,masaki2015impurity}. Unlike the CdGM modes, which are always discrete, the spectrum of chiral edge modes at the outer boundary becomes continuous in the thermodynamic limit. Note that chiral edge state on the outer boundary are unaffected by BEC.

\begin{figure}[tbp!]
  \centering
  \includegraphics[width=8cm]{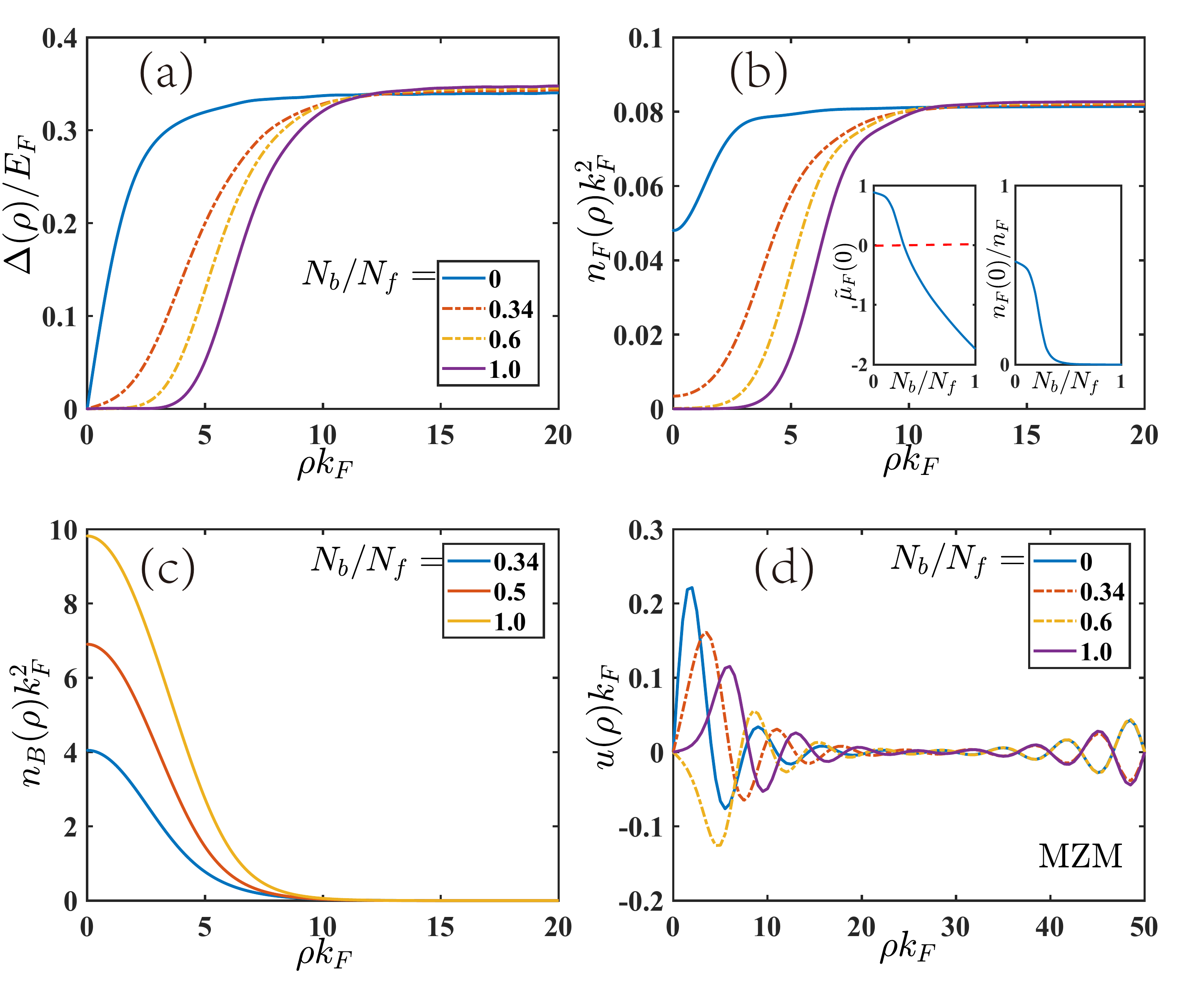}
  \caption{Spatial profiles of various quantities of a vortex core with $m=1$ and
  for different atomic-number ratios between the bosonic and fermionic components $N_b/N_f$. (a) The pairing order parameter $\Delta(\rho)$, (b) the Fermi density distribution $n_F(\rho)$, (c) the bosonic density distribution $n_B(\rho)$, and (d) the wave function of the MZM $u(\rho)$ where for MZM the $v(\rho)=u(\rho)$. Insets in (b): variations of the Fermi density $n_F(0)$ (right) and the effective chemical potential $\tilde{\mu}_{F}(0)=\mu_F-g_{BF}n_B(0)$ (left) at the center of the vortex core with increasing $N_b/N_f$. All other parameters are the same with those in Fig.~\ref{fig:fig2}.
  }
  \label{fig:fig4}
\end{figure}

More importantly, by comparing Fig.~\ref{fig:fig3}(a) and (b)(c), we observe that the number of vortex-core modes increases in the presence of BEC. Further, when the vortex core is completely depleted, as is the case in Fig.~\ref{fig:fig3}(c), the CdGM modes become fully connected to the MZM to form the branch of topological chiral edge states running around the dynamically-generated boundary between the localized BEC and the Fermi superfluid. These chiral Majorana edge states are visualized as the in-gap arrow-shaped bright stripes in Fig.~\ref{fig:fig3}(f).
%In fact, as we show in Fig.~\ref{fig:}(a), the CdGM modes adiabatically approach the MZM when the density of BEC increases.

\begin{figure*}[tbp!]
  \centering
  \includegraphics[width=13cm]{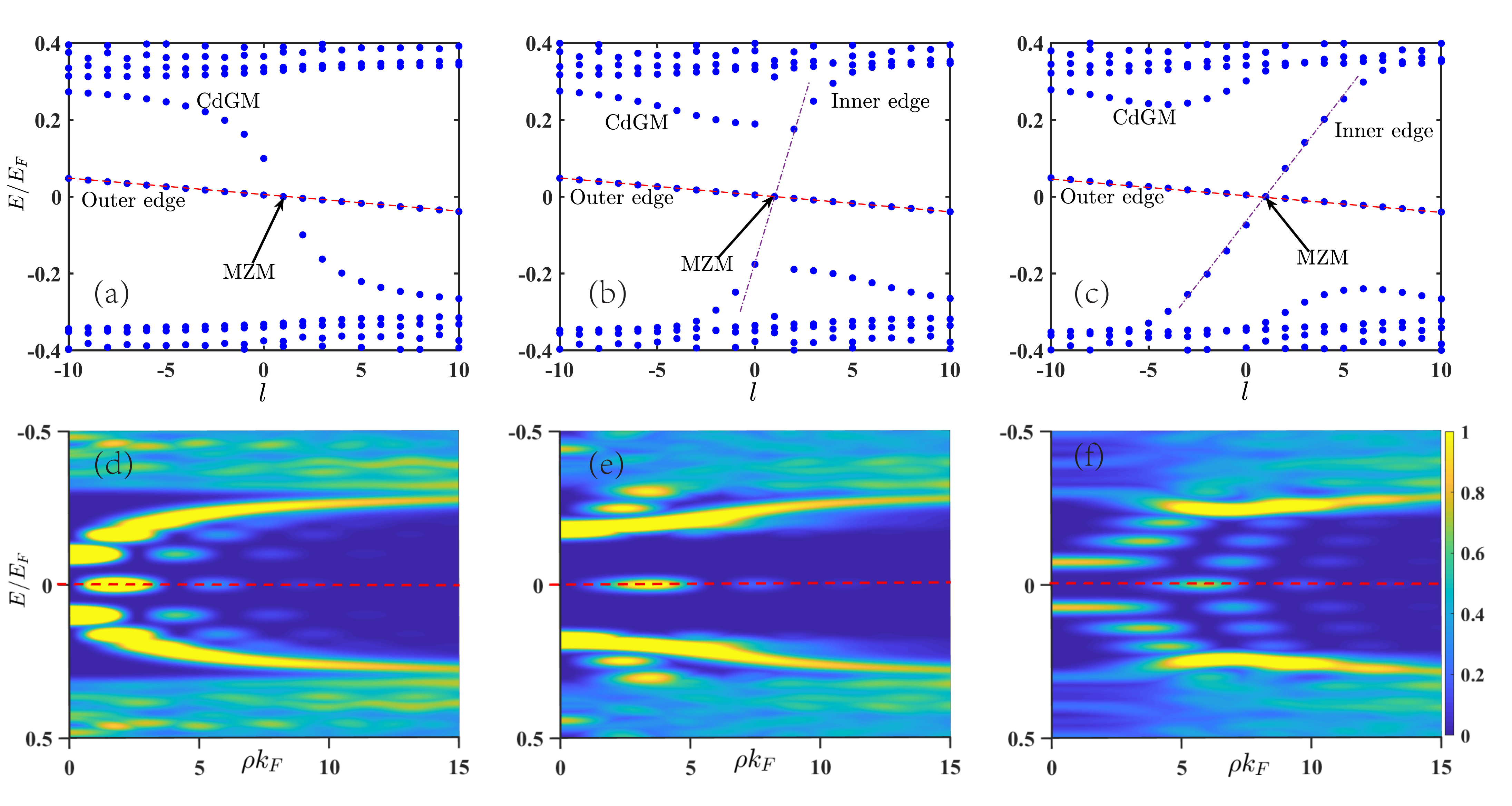}
  \caption{Quasi-particle spectra (a)(b)(c) and LDOS $D(\rho,E)$ (d)(e)(f) of a single vortex with $m=1$. (a)(d)  In the absence of BEC with $N_b/N_f=0$; (b)(e)$N_b/N_f=0.34$; (c)(f) $N_b/N_f=1$ with a fully depleted vortex core. All other parameters are the same with those in Fig.~\ref{fig:fig2}.}
  \label{fig:fig5}
\end{figure*}

Since the Fermi pairing order parameter breaks the time-reversal symmetry, topological edge states have specific chirality. As illustrated in Fig.~\ref{fig:fig1}, the running directions of edge states at the inner and outer boundaries are opposite to each other. For the vortex with $m=-1$, the occupied edge modes at the outer boundary have non-negative angular momentum and run along the boundary in the anti-clockwise direction, as schematically indicated by black arrows on the outer boundary in Fig.~\ref{fig:fig1}(a). By contrast, edge modes at the inner boundary run in a clockwise fashion along the boundary, as indicated by black arrows on the inner boundary in Fig.~\ref{fig:fig1}(a). Importantly, the chirality of Majorana edge states at the inner boundary is determined by the chirality of the Fermi superfluid, which matches the vorticity of the vortex for $m=-1$. All these features are confirmed in Fig.~\ref{fig:fig3}.

\section{Chiral Majorana edge states in a vortex with $m=1$}

We now turn to the case where the Fermi vortex features $m=1$. As illustrated in Fig.~\ref{fig:fig4}, variations of the vortex-core structure in terms of spatial profiles of the order parameter and number densities are similar to the case of $m=-1$. Here, BEC still becomes localized beyond a threshold bosonic density, where fermions are depleted from the vortex core.
However, an important difference lies in the mismatch between the chirality of the Majorana edge states on the inner boundary and the vorticity of the Fermi vortex [see Fig.~\ref{fig:fig1}(b)]. This is manifested in Fig.~\ref{fig:fig5}(a), where CdGM states traverse the gap in the same direction as chiral edge states at the outer boundary. As the BEC density increases and a dynamic boundary gradually becomes well-defined within the vortex core, a new branch of CdGM states with a matching angular momentum to the chiral superfluid develops from the bulk, which eventually forms chiral Majorana edge states running on the inner boundary [see Fig.~\ref{fig:fig5}(a)(b)(c)].

\begin{figure}[tbp]
  \centering
  \includegraphics[width=8cm]{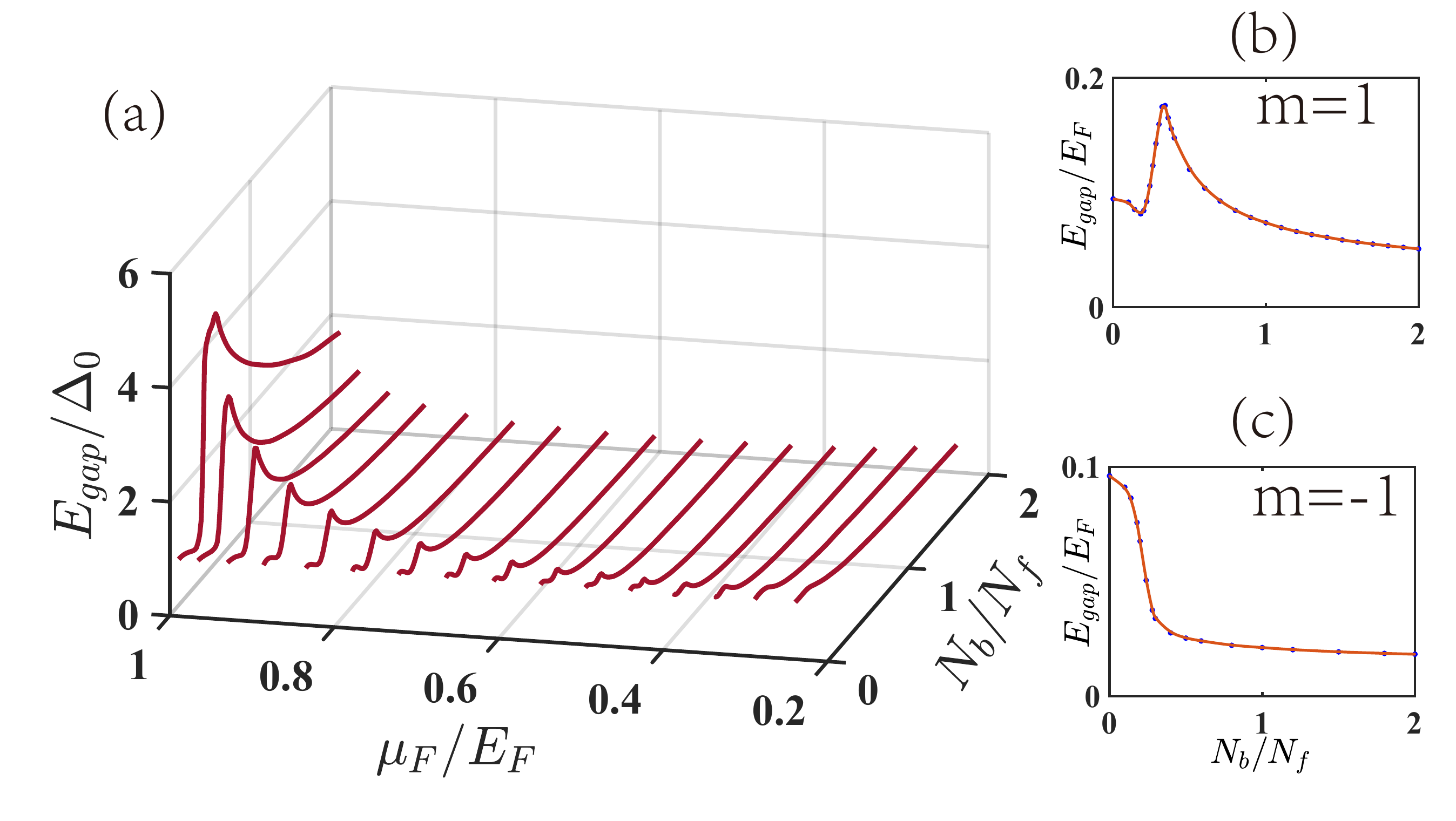}
  \caption{(a) The excitation gap of the MZM as a function of $N_b/N_f$ for different $\mu_F$ with $m=1$. (b) The excitation gap of the MZM at a fixed $\mu_F=0.9$ and with $m=1$. (c) The excitation gap of the MZM at a fixed $\mu_F=0.9$ and with $m=-1$. All other parameters are the same as Fig.~\ref{fig:fig2}.}
  \label{fig:fig6}
\end{figure}

As a result of the process discussed above, with increasing BEC density, a first-order transition takes place in the lowest excited state of the vortex core, where CdGM states with opposite angular momenta compete with each other. This gives rise to a sharp peak in the excitation gap of the MZM, as show in Fig.~\ref{fig:fig6}(a)(b), and is in contrast to the case with $m=-1$, where the excitation gap monotonically decreases [see Fig.~\ref{fig:fig6}(c)]. At the point with the largest gap, the MZM is well-separated from other core modes [Fig~\ref{fig:fig5}(e)], which should facilitate its detection. In Fig.~\ref{fig:fig6}(a), we also show the excitation gap for different chemical potentials $\mu_F$, where it is apparent that the transition is sharper and the enhancement of gap is more prominent in the weak-coupling regime.

\section{Conclusion}
In summary, we study a single vortex in mixture of $p+ip$ Fermi superfluid and a BEC. Owing to the repulsive Bose-Fermi interaction, BEC can localize at the Fermi vortex core, giving rise to a dynamically-generated boundary between the Fermi and Bose components. As a result, chiral Majorana edge states can emerge on the boundary. We study in detail how chiral Majorana edge states evolve into existence as the geometry of the topological defect at the vortex core changes with an increasing BEC density. Our work reveals that, vortices with different chirality in the $p+ip$ Fermi superfluid show great difference in response to the BEC.
Our study demonstrates an interesting example on how the geometric configurations of topological defects can be dynamically generated and controlled in a realistic system.

Our result further suggests that, in a conventional topological superconductor~\cite{Sun2016Majorana}, it is possible to add an appropriate local electric field, which should play a similar role as the BEC, generating a local interface within the core and increasing the excitation gap of the MZM.
Such a scheme could be useful for detecting and manipulating the MZMs, where adiabaticity can be facilitated by larger excitation gaps above MZMs.

\section*{Acknowledgements}
This work has been supported by the Natural Science Foundation of China (11522545) and the National Key R\&D Program (Grant Nos. 2016YFA0301700,2017YFA0304100). J.-S. P. acknowledges support from National Postdoctoral Program for Innovative Talents of China under Grant No. BX201700156.

\bibliographystyle{apsrev4-1}

%\bibliography{vortexpwave_reference}

\end{document}